\documentstyle[11pt,pasp,twoside]{article}
\input{epsf}
\def\edcomment#1{\iffalse\marginpar{\raggedright\sl#1\/}\else\relax\fi}
\reversemarginpar

\begin{document}
\title{Phase-resolved spectra of the millisecond pulsar SAX J1808.4--3658} 
\author{Marek Gierli{\'n}ski, Chris Done}
\affil{Dept. of Physics, Univ. of Durham, South Rd, Durham, DH1 3LE, UK} 
\author{Didier Barret} 
\affil{CESR, 9 av. du Col. Roche, BP 4346-31028, Toulouse, Cedex 4, France}

\begin{abstract}
  We analyze the {\it RXTE\/} observations of the April 1998 outburst
  of SAX J1808.4--3658. We show that the soft blackbody component of
  the X-ray spectrum is from the hotspot on the neutron star surface,
  while the hard component is from Comptonization of these blackbody
  photons in the plasma heated by the accretion shock. The disc cannot
  provide enough seed photons for Comptonization. It is truncated into
  an optically thin flow at $\sim$ 20--40 $R_g$ by some generic
  mechanisms common for black holes and neutron stars. Only below the
  truncation radius the flow is disrupted by the magnetic field and
  collimated onto the surface. The pulse profiles of both spectral
  components imply several constraints on the geometry of the system:
  the inclination angle, location and size of the hotspot and the
  shock.
\end{abstract}

\begin{figure}
\plotone{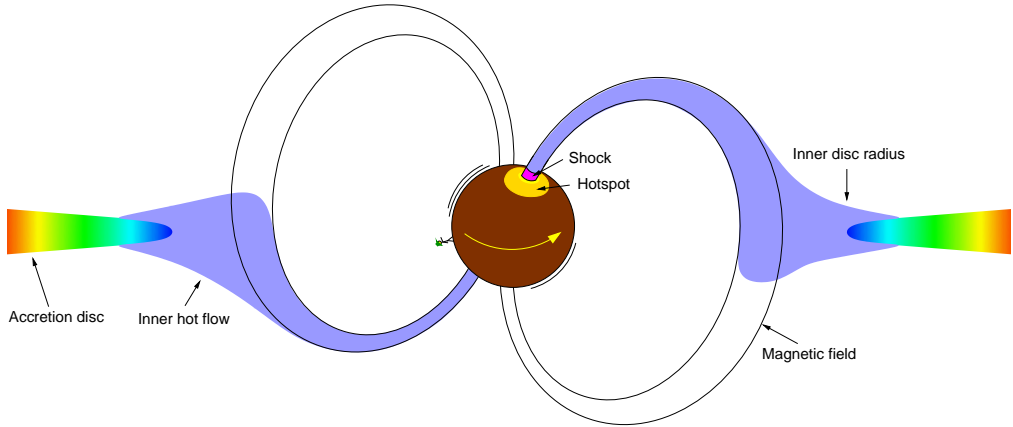}
\caption{Geometry of SAX J1808.4--3658. }
\label{fig:ns}
\end{figure}

The X-ray spectrum of the millisecond pulsar can be fitted by a
two-component model consisting of a soft blackbody and a hard
Comptonization, with weak Compton reflection. We attribute the soft
component to the hotspot on the neutron star and the hard component to
a plasma heated in the accretion shock as the material collimated by
the magnetic field impacts onto the surface. The soft and hard fluxes
are tightly correlated, so probably the hotspot is due to reprocessing
of the kinetic energy of the accretion flow and hard X-ray
irradiation. The geometry advocated here (see Figure \ref{fig:ns})
implies that the hotspot is the dominant source of seed photons for
Comptonization in the shock, while the seed photons from the disc or
self-absorbed synchrotron are negligible.

\begin{figure}
\plotone{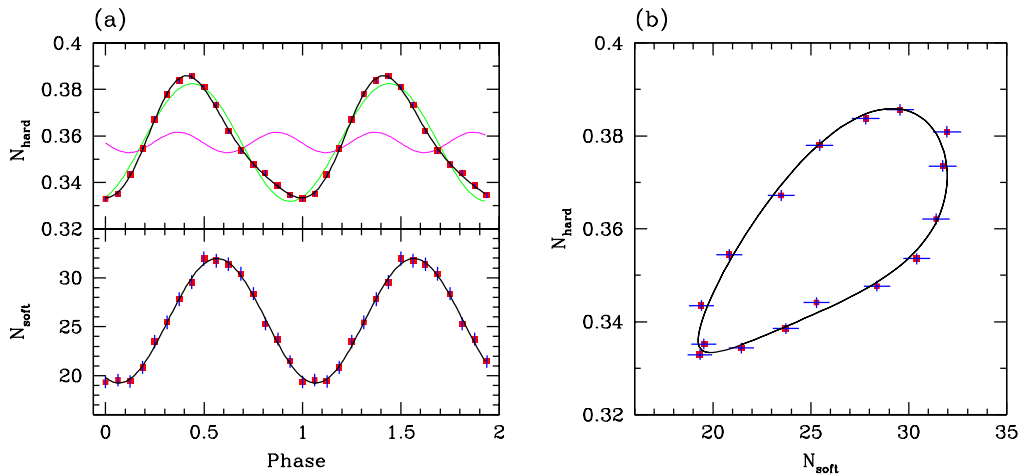}
\caption{Pulse profiles of the soft and hard spectral
  components. $N_{\rm soft}$ and $N_{\rm hard}$ are the soft and hard
  component normalizations. }
\label{fig:phase_param}
\end{figure}

In the phase-resolved spectra the two components pulse independently,
retaining their spectral shape as a function of spin phase. The
constancy of the hard X-ray spectral shape implies that the hard
X-rays are from a single emission region, rather than being the sum of
a spatially separated pulsed and unpulsed components.  The large
amplitude of variability of the hotspot (see Figure
\ref{fig:phase_param}$a$) implies large inclination of the system,
$\sim 60\deg$, {\em if\/} the angle between spin and magnetic axes is
small ($\sim 10\deg$).  Lack of secondary maxima in both profiles
implies that we do not see the second magnetic pole -- neither the
hotspot nor the shock. This, in turn, constrains the shock height to
be less than $\sim$ 10\% of the neutron star radius.  The pulse
profiles of the soft and hard components and are shifted in phase by
about $50\deg$ (Figure \ref{fig:phase_param}). The soft blackbody
component is optically thick, so its variability is dominated by its
changing projected area, while the Doppler shifts (which are maximised
$90\deg$ before the maximum in projected area) are somewhat stronger
for the translucent accreting shock. This phase shift will manifest
itself as the observed energy-dependent time lag in which soft photons
lags hard ones. A skewness of the hard component pulse profile (Figure
\ref{fig:phase_param}$a$) can be a sign of an elongated cross-section of
the shock.

The inner disc radius, $\sim$20--40 $R_g$, derived from the amount of
reflection is consistent with that derived from the relativistic disc
models for the origin of the QPO and break in the power density
spectra. The evolution of the inner radius is {\em not\/} consistent
with the magnetic field truncating the disc. Instead it seems more
likely that the inner disc radius is set by some much longer
time-scale process connected to the overall evolution of the accretion
disc. This disc truncation mechanism would then have to be generic in
all low mass accretion rate flows both in disc accreting neutron stars
and black hole systems.

%We do not report reflection from the neutron star surface.

\end{document}